\documentstyle[12pt]{article}

\author{A. B{\"o}hm\footnote {e-mail: bohm@physics.utexas.edu}, 
H. Kaldass, P. Patuleanu\footnote {Department of Physics, 
The University of Texas at Austin, Austin, Texas 78712}}
\title{\bf Hilbert Space or Gelfand Triplet - Time Symmetric or Time Asymmetric Quantum Mechanics
}
\input tcilatex

\QQQ{Created}{
September, 18, 1997
}

\QQQ{Language}{
American English
}

\begin{document}

\maketitle
\newpage
\begin{abstract}
Intrinsic microphysical irreversibility is the time asymmetry observed in
exponentially decaying states. It is described by the semigroup generated by
the Hamiltonian ${\it H}$ of the quantum physical system, not by the
semigroup generated by a Liouvillian ${\it L}$ which describes the
irreversibility due to the influence of an external reservoir or measurement
apparatus. The semigroup time evolution generated by ${\it H}$ is impossible
in the Hilbert Space (HS) theory, which allows only time symmetric boundary
conditions and an unitary group time evolution. This leads to problems with
decay probabilities in the HS theory. To overcome these and other problems
(non-existence of Dirac kets) caused by the Lebesgue integrals of the HS,
one extends the HS to a Gel'fand triplet, which contains not only Dirac
kets, but also generalized eigenvectors of the self-adjoint ${\it H}$ with
complex eigenvalues ($E_R-i\Gamma /2$) and a Breit-Wigner energy
distribution. These Gamow states $\psi ^G$ have a time asymmetric
exponential evolution. One can derive the decay probability of the Gamow
state into the decay products described by $\Lambda $ from the basic formula
of quantum mechanics ${\cal P}(t)=Tr(|\psi ^G\rangle \langle \psi ^G|\Lambda
)$, which in HS quantum mechanics is identically zero. From this result one
derives the decay rate ${\dot {{\cal {P}}}}(t)$ and all the standard
relations between ${\dot {{\cal {P}}}}(0)$, $\Gamma $ and the lifetime $\tau
_R$ used in the phenomenology of resonance scattering and decay. In the Born
approximation one obtains Dirac's Golden Rule.
\end{abstract}

\newpage
\baselineskip=24pt

\section{Extrinsic vs. Intrinsic Microphysical Irreversibility}

Irreversible time evolution of a microphysical system occurs extrinsically,
as a result of interaction with an external system such as a reservoir or a
measuring apparatus, or intrinsically as derived from the dynamics of the
system. In the Hilbert Space Quantum Mechanics, the time evolution described
by the Hamiltonian must be time reversible, leading to a widespread
conclusion that intrinsic irreversibility does not exist. Several authors,
however, noticed examples of a microphysical arrow of time. Before
discussing further these recent views, a brief exposition of extrinsic
irreversibility is given.

In the case of a system ${\cal S}$ interacting with a reservoir ${\cal R}$,
the time evolution of the density operator ${\cal \rho }(t)$ is given by the
master equation \cite{Prigogine1}, \cite{Davis}: 
\begin{equation}
\label{0.1}\frac{\partial {\cal \rho }}{\partial t}=L{\cal \rho }(t).
\end{equation}
The Liouville operator $L$ takes the form: 
\begin{equation}
\label{0.2}L{\cal \rho }(t)=-i\left[ {\it H},{\cal \rho }(t)\right] +\delta 
{\it H}{\cal \rho }(t).
\end{equation}
Without the term $\delta {\it H}{\cal \rho }(t)$, the equations (\ref{0.1})
and (\ref{0.2}) would be the von Neumann equations describing the reversible
time evolution of an (closed) isolated quantum system; their solution being
the unitary group evolution 
\begin{equation}
\label{0.3}{\cal \rho }(t)=e^{-i{\it H}t}{\cal \rho }(0)e^{i{\it H}t};\
-\infty <t<\infty .
\end{equation}
Since the Liouville operator of extrinsic irreversibility has the additional
term $\delta {\it H}{\cal \rho }(t)$, representing the effect of the
reservoir ${\cal R}$ on the system ${\cal S}$, the state does not evolve any
more according to the unitary group generated by the Hamiltonian as in (\ref
{0.3}). Instead, under certain additional conditions on the term $\delta 
{\it H}{\cal \rho }(t)$, the integration of equation (\ref{0.2}) leads to a
semigroup evolution: 
\begin{equation}
\label{0.4}{\cal \rho }(t)=\Lambda (t){\cal \rho }(0),\ \Lambda (t)=e^{Lt},\ 
\text{for }t\geq 0
\end{equation}
where $\Lambda (t)$ is the Kossakowski-Lindblad semigroup \cite{Ghirardi}, 
\cite{Antoniou}. The semigroup time evolution describes extrinsic
irreversibility because it applies to a combined system ${\cal S}\otimes 
{\cal R}$, where ${\cal S}$ does not act on ${\cal R}$, but ${\cal R}$ acts
on ${\cal S}$.

The non-quantum mechanical term $\delta {\it H}{\cal \rho }$ in the right of
equation (\ref{0.2}) does not come from the intrinsic dynamics of ${\cal S}$%
. It is an empirical term, in the sense that every reservoir ${\cal R}$ has
its own way $\delta {\it H}$ to act on ${\cal \rho }(t)$. In this paper we
shall not discuss extrinsic irreversibility, but intrinsic irreversibility.

In contrast to extrinsic irreversibility, intrinsic irreversibility is
inherent to the dynamics of the quantum system; thus even closed (isolated)
quantum systems can have irreversible time evolution. The unitary group
evolution (\ref{0.3}) is only a special case that applies to some (e.g.
stationary) but not all quantum systems. The conventional opinion was that
irreversible semigroup time evolution generated by the Hamiltonian of the
quantum system is not possible. However, some suggestions of intrinsic
irreversibility and time-asymmetry in quantum physics have been mentioned in
the past:

\begin{enumerate}
\item  According to the work of R. Peierls \cite{Garcia}, \cite{Hernandez}, 
\cite{Mondragon}, \cite{Peierls}, \cite{Peierls1} and his school,
irreversibility is connected with the choice of initial and boundary
conditions for the solutions of the Schr{\"o}dinger equation. These new
(purely outgoing) boundary conditions lead to microphysical irreversibility.

\item  T.D.Lee \cite{Lee} explained that the time reverse of a decay process
is highly improbable. The decay products have a fixed phase relationship. To
reverse this decay process would require the preparation of a state
consisting of two (or more) highly correlated incoming spherical waves with
fixed relative phase. However, it is practically impossible to build an
experimental apparatus that prepares two incoming waves with a fixed
relative phases.

\item  Ludwig \cite{Ludwig} had noticed that a state $\varphi $ must be
prepared first (at $t=0$), before an observable $|\psi (t)\rangle \langle
\psi (t)|$ can be measured in it. This implies that a detector that is to
register an observable in the state $\varphi $ must be turned on during a
time interval of positive time, i.e. at a time after the preparation
apparatus (e.g. accelerator) has been turned on. This means the observable
can only be translated to positive times, not to arbitrary negative times.
Consequently, the time evolution operator of the observable should form only
a semigroup ${\it U}_{+}(t)=e^{i{\it H}t},\ t\geq 0$. However, realizing
that a time evolution semigroup generated by the Hamiltonian was not
possible within the mathematics of the Hilbert Space, Ludwig extrapolated
this semigroup to all times $-\infty <t<\infty $.

\item  Prigogine \cite{Antoniou1}, \cite{Petrosky}, \cite{Prigogine2}, \cite
{Prigogine3} had emphasized for a long time that irreversibility is
intrinsic to the dynamics of the microsystem rather than caused by external
influences of a reservoir or a measurement apparatus. Consequently, he
demanded that irreversibility be connected with the Hamiltonian of the
quantum system.
\end{enumerate}

The most prominent example of intrinsic irreversibility is the time
evolution of resonances. Resonances cannot be described within Hilbert Space
(HS) quantum mechanics as autonomous systems. We shall show in this paper
that the same mathematics that had originally been introduced to justify the
Dirac formalism and the nuclear spectral theorem, namely the Rigged Hilbert
Space (RHS), also describes the irreversible decay of microsystems and
allows for a mathematical theory of the quantum mechanical arrow of time.

\section{Hilbert Space Idealization of Quantum Mechanics}

The first attempt to put the ideas of quantum mechanics into some
mathematical structure was achieved by Dirac \cite{Dirac}. Dirac introduced
the bras $\langle E|,\langle x|$, the kets $|E\rangle ,|x\rangle $ and an
algebra of observables generated by such operators as the Hamiltonian $H$,
the position $Q$, the momentum $P$, etc. $...$ demanding that they fulfill
the eigenvector equations: 
\begin{equation}
\label{1.1}
\begin{array}{c}
H|E\rangle =E|E\rangle  \\ 
Q|x\rangle =x|x\rangle 
\end{array}
\end{equation}
In analogy to the basis vector expansion in the 3-dimensional space ${\bf %
\vec x}=\sum\limits_{i=1}^3{\bf \vec e}_ix^i$, Dirac postulated that the
kets introduced form a complete basis system. This means that any vector $%
\phi $ can be written as: 
\begin{equation}
\label{1.2}\phi =\sum\limits_{n=1}^\infty |E_n)(E_n|\phi
)+\int\limits_0^\infty dE|E\rangle \langle E|\phi \rangle 
\end{equation}
or as:%
$$
\phi =\int\limits_{-\infty }^\infty dx|x\rangle \langle x|\phi \rangle . 
$$
In here $|E_n)$ are the eigenvectors of ${\it H}$ with discrete eigenvalues $%
E_n$ and $|E\rangle $ are the eigenvectors of the Hamiltonian with
eigenvalues $E$ from a continuous set (for which we choose ${\bf R}_{+}$).
Comparing the basis vector expansion in the 3-dimensional case with Dirac's
expansion, it is seen that the scalar products $x^i={\bf \vec e}_i\cdot {\bf %
\vec x}$ correspond to the factors $(E_n|\phi )$, $\langle E|\phi \rangle $
and $\langle x|\phi \rangle $. Consequently,these factors were understood as
scalar products measuring the components of the vector $\phi $ along the
basis vectors formed by the eigenvectors of the observable. This
interpretation is mathematically sound in the case of discrete eigenvectors $%
|E_n)$, which are elements of some Hilbert space ${\cal H}$. However, for
the continuous eigenvectors, the kets $|E\rangle $ or $|x\rangle $ are not
in ${\cal H}$ and the energy wavefunctions $\langle E|\phi \rangle $ or the
position wavefunctions $\langle x|\phi \rangle $ are not scalar products,
but generalizations thereof.

The mathematics available at that time could not encompass Dirac's
formalism. In spite of this, it became the primary calculative tool for
physicists, without having any rigorous mathematical foundation. In fact, it
was not until Schwartz's theory of distributions \cite{Schwartz} that the
Dirac's delta function became mathematically defined, and Gel'fand's theory
of RHS \cite{Gel'fand},\cite{Maurin1} that Dirac's kets $|E\rangle $ and $%
|x\rangle $ received a mathematical interpretation.

After Dirac's ideas, the first attempt at a rigorous mathematical theory for
the quantum mechanics was provided by Weyl \cite{Weyl} and von Neumann \cite
{Neumann} using the mathematics that was available at that time: the
mathematics of the Hilbert space.

A Hilbert space (H.S.) is the completion of a linear space with scalar
product $\langle \varphi |F\rangle =(\varphi |F)$, which defines the norm $%
\left\| \varphi \right\| =\sqrt{(\varphi ,\varphi )}$. A linear space ${\bf %
\Phi }$ with scalar product is incomplete if not all Cauchy sequences have
limit elements in that space. Physicists usually do not worry about the
completion of their Hilbert space, mathematicians call such spaces
pre-Hilbert spaces. The Hilbert space ${\cal H}$ is obtained by completing $%
{\bf \Phi }$, i.e. appending to ${\bf \Phi }$ all (limit elements of) Cauchy
sequences. According to the HS formulation of quantum mechanics \cite
{Neumann} there is a one-to-one correspondence between vectors in the
Hilbert space and pure physical states, and between self-adjoint operators
on the Hilbert space and observables.

The wave function $\langle x|\psi \rangle =\psi (x)$ gives the probability $%
|\psi (x)|^2\Delta x$ to detect the particle in the position interval $%
\Delta x$. The wave function $\phi (E)=\langle E|\phi \rangle $, in the
energy representation, describes the energy distribution of, e.g., a
particle beam produced by the accelerator ($|\phi (E)|^2$ represents the
energy resolution of the experimental apparatus). Physicists always
associate smooth functions with these quantities. In the Hilbert space
formulation of quantum mechanics, these wavefunctions are elements of the
space of Lebesgue square integrable functions on the real line $L^2({\bf R})$%
\cite{Maurin}. Elements of $L^2({\bf R})$ are classes of Lebesgue square
integrable functions $\{\psi (x)\}$ or $\{\phi (E)\}$ that may vary widely
on a set of Lebesgue measure zero (e.g. all rational numbers). One and the
same wave function $\phi (E)$ can be given by any function in the class $%
\{\phi (E)\}$ not only by the smooth function of this class. In addition
there are classes that do not contain a smooth function while still being
Lebesgue square integrable. This feature contradicts the physical intuition
since the wave function connected with the experimental apparatus is always
thought of as a smooth function $|\phi (E)|^2$. While it is true that the
space of smooth (infinitely differentiable and rapidly decreasing)
functions, ${\cal S}$, is a dense subset of $L^2({\bf R})$, ${\cal S}$ is
not complete in the norm defined by the scalar product. Insisting on a
complete topological vector space, mathematicians chose the Hilbert space $%
L^2$ for quantum mechanics, because the space ${\cal S}$, whose completion
is defined not by one norm but by a countable number of norms $\left\| \cdot
\right\| _1$, $\left\| \cdot \right\| _2$, ... $\left\| \cdot \right\| _p$,
..., did not exist at that time.

To each smooth function $\psi ^{smooth}(x)\in {\cal S}$ one can always find
a class of Lebesgue square integrable functions $\{\psi (x)\}$ to which $%
\psi ^{smooth}(x)$ belongs, i.e. ${\cal S}\subset L^2$ but not vice versa:
there are classes of Lebesgue square integrable functions $\{\chi (x)\}\in
L^2$ which do not contain any smooth function, because the space of smooth
functions ${\cal S}$ is not complete with respect to the norm $\left\| \psi
\right\| ^2=\int |\psi (x)|^2dx$. If experiments provide only smooth wave
functions (they measure only in a finite set of intervals and interpolate
smoothly between these intervals) then the space ${\cal S}$ should be
sufficient for all states $\psi $ connected with experimental apparatuses.
Thus the complete Hilbert space $L^2$ is too big.

On the other hand, the HS does not contain Dirac's kets and bras, because
the eigenstates of the continuous spectrum are not in the HS and certainly
the HS does not posses a complete basis system in the sense of (\ref{1.2}).
However these kets, e.g. the scattering ''states'' $|\vec p\ \rangle $ with
momentum eigenvalue $\vec p$, have been very useful for scattering
experiments. Thus the Hilbert Space $L^2$ is also too small, since it does
not contain these scattering states. We shall therefore seek a formulation
which will overcome these problems. As an unexpected bonus, this new
formulation will also contain vectors that represent exponentially decaying
states and will describe irreversible processes like quantum decays.

\section{Problems with Quantum Decay}

Most practical computations in physics do not rely on the completeness
property of the Hilbert space and use only the pre-Hilbert space. However,
investigating general properties of decay, the full mathematical structure
of the Hilbert space has led to general results which are not desirable for
a theory of decaying phenomena.

The first property of the HS formulation to note is that symmetry groups,
like the Galileo group and the Poincar{\'e} group, are represented by
unitary operators. In the Hilbert space, for a system described by a
Hamiltonian ${\it H}$, the time evolution is given by an unitary group: 
\begin{equation}
\label{2.1} 
\begin{array}{ccc}
\phi (t)=e^{-i{\it H}t}\phi (0)={\it U}^{\dagger }(t)\phi (0)\  &  & \
-\infty <t<\infty 
\end{array}
\end{equation}
and is reversible. Therefore, any physical state in the HS can be evolved to
any instant in the past and in the future. This defies the physical
intuition regarding the evolution of resonance states backwards to instances
before their production. Microphysical irreversibility, as exemplified by
the time evolution of resonances or decaying states, is ruled out by this
unitary group evolution.

Another important feature of the HS quantum mechanics relevant to decay
processes is that no exponential decay law can be obtained within the
framework of Lebesgue square integrable functions \cite{Khalfin}. This
result deals with the time behavior of the survival probability of a given
state $\phi (0)$. The survival probability is the probability to find, at
any given time $t$, the state $\phi (0)$ in the time evolved state $\phi
(t)=e^{-i{\it H}t}\phi (0)$: 
\begin{equation}
\label{2.2}{\cal P}_S(t)=|\langle \phi (0)|e^{-i{\it H}t}|\phi (0)\rangle |^2
\end{equation}
Khalfin's theorem states that there is no HS vector $\phi (0)$ for which (%
\ref{2.2}) obeys the exponential law. This ''deviation from the exponential
law'' has at least so far not been confirmed experimentally. But since the
mathematics of the Hilbert space cannot predict the magnitude of such a
deviation, it will always remain an untestable mathematical prediction
(because the deviation could always be smaller than the available
experimental accuracy). Therefore the more practical attitude is to find a
mathematical formulation that upholds the empirical exponential law for the
resonance state and attribute whatever deviations may be observed in the
future to the admixture of some background. The description of exponential
decay can be best accomplished, as envisioned by Gamow \cite{Gamow}, if one
uses an eigenvector $\psi ^G\equiv |E_R-i\Gamma /2\rangle $ of the
self-adjoint Hamiltonian that has a complex eigenvalue $E_R-i\Gamma /2$, 
\begin{equation}
\label{2.3}{\it H}|E_R-i\Gamma /2\rangle =(E_R-i\Gamma /2)|E_R-i\Gamma
/2\rangle .
\end{equation}
As mentioned above such vectors do not exist in the HS, but they exist in
the RHS.

The strongest evidence for the inadequacy of the HS in the description of
decay phenomena is that the decay probability is zero for all time, if it is
zero on a finite time interval. The decay probability is the probability for
the transition from a state $\phi (t)=e^{-i{\it H}t}\phi (0)={\it U}%
^{\dagger }(t)\phi (0)$ into the decay products described by the subspace $%
\Lambda {\cal H}\subset {\cal H}$, where $\Lambda $ is the projection
operator on the subspace of non-interacting decay products. This decay
probability is given by: 
\begin{equation}
\label{2.4}{\cal P}_\Lambda (t)=Tr(\Lambda |\phi (t)\rangle \langle \phi
(t)|) 
\end{equation}
Since the decay of a prepared quasi-stationary state starts at a finite time 
$t>t_2>-\infty $, the probability to detect the observable $\Lambda $ in the
state $|\phi (t)\rangle $ during a time interval $-\infty \leq t_1<t_2$
should be zero. In other words: 
\begin{equation}
\label{2.5}\int\limits_{t_1}^{t_2}{\cal P}_\Lambda
(t)dt=\int\limits_{t_1}^{t_2}\langle \phi (t)|\Lambda |\phi (t)\rangle dt=0 
\end{equation}
Then it follows from Hegerfeldt's theorem \cite{Hegerfeldt}, if $\phi
(t)=e^{-i{\it H}t}\phi (0)\in {\cal H}$, with the Hamiltonian ${\it H}$
being self-adjoint and semibounded, that ${\cal P}_\Lambda (t)=0$ for
(almost) all $t$ (future $t>t_2$ and past $t<t_1$). This means that,
according to the Hilbert space formulation, there can be no decay of a state 
$\phi (t)$ which has been produced at any finite time $t_2(\neq -\infty )$.

Summarizing, in the HS quantum mechanics the time evolution is reversible,
there is no exponential decay law and the decay probability is identically
zero. These are the underlying reasons for which in practical calculations
resonances could not be described in the Hilbert space. Instead, resonances
were successfully described by ''effective theories'' as eigenvectors of
some finite dimensional complex Hamiltonians \cite{Lee-Oehme}. As will be
seen, the RHS provides a mathematical theory which will overcome the
problems of the HS theory. In addition the RHS will contain a finite
dimensional subspace in which the effective theories reappear as truncations
of a complex basis vector expansion.

\section{Rigging the Hilbert Space into a Gel'fand Triplet}

A Rigged Hilbert Space is constructed on the structure of a linear space, $%
{\bf {\Psi }}$, with a scalar product $\langle \phi |F\rangle =(|\psi
\rangle ,|F\rangle )$, through the completion of the space ${\bf {\Psi }}$
with respect to different topologies. The completion of ${\bf {\Psi }}$,
with respect to a topology $\tau $, contains all the limit points of the
Cauchy sequences in the respective topology. The Hilbert space ${\cal H}$ is
the completion of ${\bf {\Psi }}$ with respect to the norm topology $\tau _{%
{\cal H}}$. The space ${\bf {\Phi }}$ is defined as the completion of ${\bf {%
\Psi }}$ with respect to a topology $\tau _{{\bf {\Phi }}}$, stronger than
the norm topology. Since ${\bf {\Phi }}$ and ${\cal H}$ are the completions
of the same space ${\bf {\Psi }}$, ${\bf {\Phi }}$ will be dense in ${\cal H}
$, with respect to the topology of the Hilbert space. The topology $\tau _{%
{\bf {\Phi }}}$ of the space ${\bf {\Phi }}$ is given by an infinite number
of norms chosen such that the algebra of observables in the space ${\bf {%
\Phi }}$ becomes an algebra of continuous operators. ${\bf \Phi }^{\times }$
(and ${\cal H}^{\times }$) denote the space of continuous antilinear
functionals over the space ${\bf {\Phi }}$ (and ${\cal H}$). These
antilinear functionals $F(\phi )$ are denoted as $F(\phi )=\langle \phi
|F\rangle $ (or $F(h)=\langle h|F\rangle $) and are defined over the set $%
\phi \in {\bf {\Phi }}$ (or $h\in {\cal H}$). There are more functionals $%
|F\rangle $ in ${\bf {\Phi }}^{\times }$ than $|F\rangle $ in ${\cal H}%
^{\times }$, and since ${\cal H}^{\times }={\cal H}$ (Frechet-Riesz
theorem), one has constructed a triplet of spaces, called {\it Gel'fand
triplet} or {\it RHS }\cite{Gel'fand}, \cite{Maurin1}. 
\begin{equation}
\label{1}{\bf \Phi \subset }{\cal H=H}^{\times }\subset {\bf \Phi }^{\times }
\end{equation}
The space ${\bf \Phi }^{\times }$ is an extension of the Hilbert space $%
{\cal H}$. The topology $\tau _{{\bf {\Phi }}}$ can be chosen such that $%
{\bf \Phi }^{\times }$, unlike the Hilbert Space ${\cal {H}}$, contains
''eigenvectors'' of a self adjoint operator with eigenvalues belonging to
the continuous spectrum, e.g. the Dirac kets. In addition ${\bf \Phi }%
^{\times }$ contains ''eigenvectors'' of self adjoint operators with complex
eigenvalues. The Dirac brac-ket $\langle \phi |F\rangle $ is just an
extension of the scalar product $(\phi ,F)$ in ${\cal H}$.

In a scattering experiment one defines two sets of Rigged Hilbert Spaces one
for the prepared in-states, ${\bf {\Phi }}_{-}$, and the other for the
observed (detected) out-states, ${\bf {\Phi }}_{+}$: 
\begin{equation}
{{\bf \Phi }_{-}\subset }{\cal H=H}^{\times }\subset {\bf \Phi }_{-}^{\times
}
\end{equation}
and 
$$
{\bf \Phi }_{+}{\subset }{\cal H=H}^{\times }\subset {\bf \Phi }_{+}^{\times
} 
$$
Where ${\bf \Phi }={\bf \Phi }_{-}+{\bf \Phi }_{+}$ and ${\bf \Phi }_{-}\cap 
{\bf \Phi }_{+}\neq {0}$. A vector $\phi ^{+}\in {\bf \Phi }_{-}$ is what is
prepared as $\phi ^{in}$ outside the interaction region, while a vector $%
\psi ^{-}\in {\bf \Phi }_{+}$ is what is registered as $\psi ^{out}$ outside
the interaction region.

The topology in the space ${\bf \Phi }$, and equivalently in the spaces $%
{\bf \Phi }_{+}$ and ${\bf \Phi }_{-}$, is always chosen in such a way that
the operators representing the observables are continuous (so bounded)
operators on ${\bf {\Phi }}$ (with respect to the topology $\tau _{{\bf {%
\Phi }}}$). In the Hilbert space the observables cannot be represented by
continuous operators (with respect to $\tau _{{\cal {H}}}$). For example:\
in the Hilbert space, if the operators $P$ and $Q$ fulfill the Heisenberg
commutation relation, then they can not be both continuous operators in $%
{\cal H}$ and in the standard representation neither $P$ nor $Q$ are
continuous operators. As ${\bf {\Phi }}$ is a dense subspace of the Hilbert
space, all the operators that are used in the Hilbert space can be redefined
in ${\bf {\Phi }}$ as restrictions to the space ${\bf {\Phi }}$. For each $%
\tau _{{\bf {\Phi }}}$-continuous operator $A$ on ${\bf {\Phi }}$ one can
define its conjugate operator $A^{\times }$, as an extension of the HS
adjoint operator $A^{\dagger }$: 
\begin{equation}
\label{3}\langle A\phi |F\rangle =\langle \phi |A^{\times }|F\rangle ,\text{%
for all }\phi \in {\bf \Phi },F\in {\bf \Phi }^{\times }.
\end{equation}
As a result, we obtain a triplet: 
\begin{equation}
\label{2}A^{\dagger }\mid _{{\bf \Phi }}\subset A^{\dagger }\subset
A^{\times }.
\end{equation}
It should be noted that the conjugate operator $A^{\times }$ can only be
defined for a $\tau _{{\bf \Phi }}$-continuous operator $A$, and,
consequently, is a continuous operator on ${\bf \Phi }^{\times }$. The
generalized eigenvector $|F\rangle $ of a continuous operator $A$ is defined
by the following relation:

\begin{equation}
\label{4}\langle A\phi |F\rangle =\langle \phi |A^{\times }|F\rangle =\omega
\langle \phi |F\rangle ,\text{ for all }\phi \in {\bf \Phi } 
\end{equation}
Ignoring the arbitrary vectors $\phi $, this is often also written as:

\begin{equation}
\label{5}A^{\times }|F\rangle =\omega |F\rangle ,
\end{equation}
or as%
$$
A|F\rangle =\omega |F\rangle , 
$$
if $A$ is essentially self adjoint. This method makes it possible to
describe ''eigenstates'' that can not exist in the Hilbert space. Some of
the generalized eigenvectors are going to be the ordinary eigenvectors of
the essentially self-adjoint operator in the Hilbert space. But not all
generalized eigenvectors are elements of the Hilbert space. In particular,
the Dirac kets, that describe the scattering states, are generalized
eigenvectors with eigenvalues belonging to the continuous spectrum and are
not in ${\cal H}$. The Gamow vectors, that describe the states with an
irreversible time evolution, are also generalized eigenvectors which are not
in ${\cal H}$, but their complex eigenvalues do not belong to the Hilbert
space spectrum of the Hamiltonian. The choice of ${\bf \Phi }$, given by the
choice of the topology $\tau _{{\bf {\Phi }}}$, determines which set of
generalized eigenvectors is possible for a given operator $A$ in ${\cal H}$.
This choice of the spaces ${\bf \Phi }$, ${\bf \Phi }_{+}$ and ${\bf \Phi }%
_{-}$ is made using physical arguments related to causality and initial and
boundary conditions. The initial conditions are determined by the setup of
the experiment. In the RHS formulation of the quantum mechanics one uses the
same dynamical equations as in the Hilbert space formalism, while the
initial (boundary) conditions are different from the HS boundary conditions.
The space ${\cal H}$ in the HS formalism describes all physical systems, for
each particular system only a dense subspace is used for practical
calculations. The spaces ${\bf \Phi }$, ${\bf \Phi }_{+}$ and ${\bf \Phi }%
_{-}$ are specific for the particular physical system under consideration.

\section{Gamow Vectors and Their Properties}

The Rigged Hilbert Space was developed in order to accommodate Dirac's kets
and bras into a consistent mathematical structure, but the structure,
created for Dirac's formalism, provided a mathematical description for the
states with an irreversible time evolution too. In the Hilbert space, an
irreversible process is possible only for an open system under the influence
of an external reservoir. There are no vectors in ${\cal H}$ which can
represent isolated microphysical states that can evolve irreversibly in
time. In the RHS, decaying states which are described by the Gamow vectors $%
|z_R^{-}\rangle \equiv |E_R-i\Gamma /2^{-}\rangle $, evolve irreversibly in
time by a semigroup generated by the Hamiltonian.

The following are the properties of Gamow vectors describing decaying states:

\begin{enumerate}
\item  They are generalized eigenvectors of the Hamiltonian associated with
the complex eigenvalue $E_R-i\Gamma /2$ (where $E_R$ and $\Gamma $ were
interpreted as the resonance energy and the width of the resonance
respectively); i.e. the following equation holds: 
\begin{equation}
\label{71}{\it H}^{\times }|z_R^{-}\rangle =\left( E_R-i\Gamma /2\right)
|z_R^{-}\rangle 
\end{equation}
as functional equation over all $\psi ^{-}\in {\bf \Phi }_{+}$ (in the sense
of (\ref{5})).

\item  They are derived as functionals from the resonance pole term at $%
z_R=E_R-i\Gamma /2$ in the second sheet of the analytically continued $S$%
-matrix.

\item  They have a Breit-Wigner energy distribution 
\begin{equation}
\label{8}\langle ^{-}E|\psi ^G\rangle =i\sqrt{\Gamma /2\pi }\frac 1{E-\left(
E_R-i\Gamma /2\right) },-\infty _{II}<E<+\infty 
\end{equation}
(where the negative values of $E$ are in the second Riemann sheet of the $S$%
-matrix).

\item  They are members of a basis system (like the Dirac kets $|E\rangle $%
), i.e.,every prepared state vector $\phi ^{+}\in {\bf \Phi _{-}}$ can be
expanded as \cite{JMP}: 
\begin{equation}
\label{9}\phi ^{+}=\sum_{n=1}^\infty |E_n)(E_n|\phi
^{+})+\sum\limits_{i=1}^N|\psi _i^G\rangle \langle \psi _i^G|\phi
^{+})+\int_0^{-\infty _{II}}dE|E^{+}\rangle \langle ^{+}E|\phi ^{+}\rangle 
\end{equation}
(where $-\infty _{II}$ indicates that the integration along the negative
real axis is in the second Riemann sheet). In contrast, the Dirac basis
system expansion (Nuclear Spectral Theorem of the RHS) is given by : 
\begin{equation}
\label{10}\phi ^{+}=\sum_{n=1}^\infty |E_n)(E_n|\phi ^{+})+\int_0^{+\infty
}dE|E^{+}\rangle \langle ^{+}E|\phi ^{+}\rangle 
\end{equation}
In here:

\begin{itemize}
\item  $|E_n),\ n=1,2,...,\infty $ are the stable eigenstates (bound state
poles),

\item  $|\psi _i^G\rangle =|E_i-i\Gamma _i/2^{-}\rangle \sqrt{2\pi \Gamma _i}
$ are the $N$ decaying(Gamow) states (resonance poles),

\item  $|E^{+}\rangle ,\ 0\leq E<\infty $ (Hilbert space spectrum) are the
Dirac scattering states,

\item  $|E^{+}\rangle ,\ -\infty _{II}<E\leq 0$ are the latter's analytic
continuation to the negative real axis on the second sheet.
\end{itemize}

The important feature of the so-called complex spectral resolution (\ref{9})
is that the resonance states $\psi _i^G$ appear on the same footing as the
bound states $|E_n)$. But together with the bound states they do not form a
complete system, there is in addition a ''background term''.

\item  The time evolution, in general, is given by a semigroup generated by
the Hamiltonian for $t\geq 0$ (a corresponding semigroup with $t\leq 0$
applies to the exponentially growing Gamow vector ${\tilde \psi }^G\in {\bf %
\Phi }_{+}^{\times }$, which is associated with the $S$-matrix pole at $%
z_R=E_R+i\Gamma /2$ )\cite{Bohm}. The unitary time evolution group applies
only to the common Hilbert subspace of ${\bf \Phi }_{+}^{\times }$ and ${\bf %
\Phi }_{-}^{\times }$. The time evolution of the decaying Gamow state in
particular is given by an exponential law: 
\begin{equation}
\label{11}e^{-i{\it H}^{\times }t}|\psi ^G\rangle \langle \psi ^G|e^{i{\it H}%
t}=e^{-i(E_R-i\Gamma /2)t}|\psi ^G\rangle \langle \psi ^G|e^{i(E_R+i\Gamma
/2)t}=e^{-\Gamma t}|\psi ^G\rangle \langle \psi ^G|
\end{equation}
for $t\geq 0$ only. This is understood as a functional equation over the
space of $\psi ^{-}\in {\bf \Phi }_{+}$. This time evolution is irreversible
because $e^{-i{\it H}^{\times }t}$ ($t\geq 0$) is a semigroup.
\end{enumerate}

\section{Decay Probability and Decay Rate in RHS}

In the RHS, the description of irreversible processes becomes possible.
Decaying states are described by Gamow vectors. A process in which a
microphysical state evolves in time and decomposes into a set of decay
products will be described as the transition of a Gamow vector $\psi ^G$
into a set of interaction free decay products.

In a decay experiment, the decaying state and the set of detected decay
products are described by different Hamiltonians. While the detected states
evolve in time according to the free Hamiltonian $H_0$ (since they are
supposed to be detected far away from the interaction zone), the decaying
state is a generalized eigenvector of the exact Hamiltonian $H=H_0+V$, where 
$V$ is the interaction responsible for the decay. The eigenkets of the free
Hamiltonian $|E,b\rangle $ are assumed to be mapped into the eigenkets of
the exact Hamiltonian $|E,b^{-}\rangle $ by the Lippmann-Schwinger equation 
\begin{equation}
\label{111}|E,b^{-}\rangle =|E,b\rangle +\frac 1{E-H-i\epsilon }V|E,b\rangle
. 
\end{equation}

Examples of decay processes are: the radiative decay of an excited atom into
its ground state ($A^{*}\rightarrow A+\gamma $) with the emission of a
photon or the decay of a $K$-meson ($K^0\rightarrow \pi ^{+}\pi ^{-}$) into
two pions.

The decay rate ${\dot {{\cal {P}}}}(t)$ of the $\psi ^G(t)$ into the final
non-interacting decay products can be calculated as a function of time and
leads to an exact Golden Rule (with the natural line width given by a
Breit-Wigner(\ref{8})). In the Born approximation the Gamow vector $\psi ^G$
goes into $f^d$ ($\psi ^G\rightarrow f^d$, which is an eigenvector of ${\it H%
}_0={\it H}-V$; $E_R\rightarrow E_d$ and $\Gamma /E_R\rightarrow 0$) and the
decay rate goes into Fermi's Golden Rule.

The time evolution of the ''pure Gamow state'' with resonance parameters ($%
E_R,\Gamma $), initially described by the statistical operator $W(0)=|\psi
^G\rangle \langle \psi ^G|$, is given, according to (\ref{11}), by 
\begin{equation}
\label{12}W(t)=e^{-i{\it {H}}t}W(0)e^{i{\it {H}}t}=e^{-\Gamma t}W(0),\ \text{
for }t\geq 0.
\end{equation}
This is mathematically defined only as a functional over $\psi ^{-}\in {\bf %
\Phi }_{+}$, where ${\bf \Phi }_{+}$ is defined as the space connected with
the decay products (''out-states''). This means that only $\langle \psi
^{-}|W(t)|\psi ^{-}\rangle $ makes sense mathematically.

The interaction free decay products are described by the projection operator
, $\Lambda $, onto the space of the physical states of all the
non-interacting decay products: 
\begin{equation}
\label{13}\Lambda =\dint dE\dsum\limits_b|E,b\rangle \langle E,b|
\end{equation}
where $|E,b\rangle $ are the eigenvectors of the free Hamiltonian 
\begin{equation}
\label{14}{\it H}_0|E,b\rangle =E|E,b\rangle 
\end{equation}
and '$b$' stands for all the possible labels of these eigenvectors. If the
system is described by a complete set of commuting operators $B_1$, $B_2$,
... $B_N$, then $b$ will be given by the set $b_1$, $b_2$, ... $b_N$ of
quantum numbers labeling the degeneracy of the energy $E$ (the $b$s can be
the quantum numbers for the orbital angular momentum, the photon
polarization $\lambda _\gamma $, some other intrinsic quantum numbers like
charges or channel labels, or the momentum directions ($\theta _k,\varphi _k$%
) of the decay products). We will use the index '$b$' for the whole set of
quantum numbers in order to simplify the formulas since the choice of these
labels will not change the results. For example in (\ref{13}), for the
process $A^{*}\rightarrow A+\gamma $:%
$$
|E,b\rangle =|E,\theta _k,\phi _k,\lambda _\gamma ...\rangle  
$$
and%
$$
\sum_b=\sum_{\lambda _\gamma }\int dcos\theta _kd\phi _k\sum_{...}, 
$$
where '...' stands for the quantum numbers of the atomic states.

The decay probability is the expectation value of the operator $\Lambda $,
for the interaction free decay products, in the state $W(t)$ of the decaying
state. Therefore it is given by the general formula for the expectation
value of an observable $\Lambda $ in a state $W(t)$:
\begin{equation}
\label{16}{\cal P}(t)=Tr(\Lambda W(t))
\end{equation}

As explained in section 3, in the Hilbert space one can prove (with the only
assumption that ${\it H}$ is self-adjoint and semibounded, which must always
be the case (stability of matter)) that ${\cal P}(t)$ is either identically
zero for all times or it has been already different from zero in a time
interval starting at $t=-\infty $. This means that in the HS one predicts no
decay for any state that has been prepared at a particular time $t_0$ ($\neq
\infty $). In the RHS one can derive from (\ref{16}), with $W(0)$ given by $%
|\psi ^G\rangle \langle \psi ^G|$ and $\Lambda $ by (\ref{13}), that 
\begin{equation}
\label{17}{\cal P}(t)=1-e^{-\Gamma t}\int dE\sum_b|\langle E,b|V|\psi
^G\rangle |^2\frac 1{(E-E_R)^2+(\Gamma /2)^2},\ \text{ for }t\geq 0. 
\end{equation}
In this derivation one uses (\ref{71}), (\ref{12}) and the
Lippmann-Schwinger equation (\ref{111}) and one chooses as boundary
conditions ${\cal {P}}(t=\infty )=1$ (meaning that after a long enough time
all the decay products have decayed and their decay products have been
measured) and ${\cal {P}}(t=0)=0$ (so that no decay product is measured
before the preparation of the decaying state is completed at time $t=t_0=0$).

An exact Golden Rule for the decay rate is obtained by taking the time
derivative of the transition probability, ${\cal {P}}$, given in (\ref{17}): 
\begin{equation}
\label{18}{\dot {{\cal {P}}}}(t)=2\pi e^{-\Gamma t}\int dE\sum_b|\langle E,b|V|\psi
^G\rangle |^2\frac{\Gamma /2\pi }{(E-E_R)^2+(\Gamma /2)^2}
\end{equation}
The decay rate ${\dot {{\cal {P}}}}(t)$ has a Breit-Wigner distribution, whose width is 
$\Gamma $. This is an exact formula from which one can obtain, in the Born
approximation, Fermi's Golden Rule if one inserts for the state $\psi ^G$
the non-interacting state $f^d$ with $H_0f^d=E_df^d$. The Born approximation
is defined by:%
$$
\langle E,b|V|\psi ^G\rangle \rightarrow \langle E,b|V|f^d\rangle  
$$
$$
E_R\rightarrow E_d 
$$
\begin{equation}
\label{20}\Gamma /2E_R\rightarrow 0
\end{equation}
$$
\frac{\Gamma /2\pi }{(E-E_R)^2+(\Gamma /2)^2}\rightarrow \delta (E-E_R) 
$$
In this approximation (\ref{20}), the initial decay rate is obtained from (%
\ref{18}) as:
\begin{equation}
\label{21}{\dot {{\cal {P}}}}(0)=2\pi \int dE\sum |\langle E,b|V|f^d\rangle |^2\delta
(E-E_R)
\end{equation}
This is the standard Golden Rule for the transition from an excited
non-interacting state $f^d$, into the set of all non-interacting decay
products.

On the other hand, using the condition ${\cal P}(0)=0$ one obtains from (\ref
{17}) in the limit (\ref{20}):
\begin{equation}
\label{22}\Gamma =2\pi \int dE\sum |\langle E,b|V|f^d\rangle |^2\delta
(E-E_R).
\end{equation}
Comparing this with (\ref{21}), one obtains that 
\begin{equation}
\label{23}{\dot {{\cal {P}}}}(0)=\Gamma 
\end{equation}
From the exponential decay law in (\ref{17}), for the survival probability $%
1-{\cal P}(t)=e^{-\Gamma t}$ or from the exponential law in (\ref{18}) for
the decay rate, one obtains that 
\begin{equation}
\label{24}\Gamma =\frac 1{\tau _R},
\end{equation}
where $\tau _R$ is the lifetime of the resonance state. The results (\ref{21}%
), (\ref{23}) and (\ref{24}) and the identification of $\Gamma $ with the
imaginary part of the resonance pole position $z_R$ of the analytically
continued $S$-matrix are the standard relations used in the analysis of
resonance scattering and decay phenomena. They have been justified by
various more or less heuristic arguments, in particular also by making use
of the exponential law for the survival probability. But they have not been
derived from the basic formula (\ref{16}) for the probabilities in quantum
mechanics, and they could not have been derived from this formula because,
applied to the probabilities of decay, this formula is identically zero in
the HS formulation \cite{Hegerfeldt}, as mentioned in section 3. The
quantity that had been missing from the HS formulation, and which is needed
to provide the theoretical link between these important empirical formulas
and the basic formula for probabilities (\ref{16}), is the Gamow vector.
Gamow vectors allow the description of resonances as elementary particles in
very much the same way as stable particles are described, either as poles of
the $S$-matrix or as energy eigenstates, only that for the Gamow states the
energy is the complex number $E_R-i\Gamma /2$ and this requires the
mathematics of the RHS.

\end{document}